\documentclass[twocolumn,
superscriptaddress,
prl,
showpacs,
preprintnumbers,
amsmath,amssymb]{revtex4}

\usepackage{graphicx,epsf,epsfig,ulem}
\usepackage{epsfig}
\usepackage{bm}
\usepackage{subfigure}
\usepackage{color}

\begin{document}

\title{Voltage controlled exchange energies of a two electron silicon double quantum dot with and without charge defects in the dielectric} 

\author{Rajib Rahman}
\affiliation{Sandia National Laboratories, Albuquerque, NM 87185, USA}

\author{Erik Nielsen}
\affiliation{Sandia National Laboratories, Albuquerque, NM 87185, USA}

\author{Richard P. Muller}
\affiliation{Sandia National Laboratories, Albuquerque, NM 87185, USA}

\author{Malcolm S. Carroll}
\affiliation{Sandia National Laboratories, Albuquerque, NM 87185, USA}

\date{\today}
\begin{abstract}
Quantum dots are artificial atoms used for a multitude of purposes. Charge defects are commonly present and can significantly perturb the designed energy spectrum and purpose of the dots. Voltage controlled exchange energy in silicon double quantum dots (DQD) represents a system that is very sensitive to charge position and is of interest for quantum computing. We calculate the energy spectrum of the silicon double quantum dot system using a full configuration interaction that uses tight binding single particle wavefunctions.  This approach allows us to analyze atomic scale charge perturbations of the DQD while accounting for the details of the complex momentum space physics of silicon (i.e., valley and valley-orbit physics). We analyze how the energy levels and exchange curves for a DQD are affected by nearby charge defects at various positions relative to the dot, which are consistent with defects expected in the metal-oxide-semiconductor system.    
\end{abstract}


\maketitle 

Electrostatically defined quantum dots (QDs) in semiconductors are one of the most promising systems for realizing a scalable quantum computer \cite{Loss}. 
These systems provide the opportunity to engineer and control quantum mechanical properties through conventional electronics \cite{Levy}, and all elements of a qubit have been demonstrated in a GaAs double quantum dot (DQD) \cite{Petta}. The presence of nuclear spins in GaAs, however, limit its spin coherence times to order of nanoseconds \cite{Petta, Taylor} without any quantum control. Silicon, on the other hand, can have exceptionally long spin coherence times, often on the order of milliseconds to seconds \cite{Tryshkin2, Witzel}, and is therefore a very promising host material for solid-state qubits. 



After a decade of extensive research, experiments are beginning to build and measure few electron silicon QDs repeatedly \cite{Eriksson, Simmons2, Dzurak, HRL, House1, Lu}. The recent demonstration of single-shot spin readout in silicon also marks a significant milestone towards building a functional silicon qubit \cite{Prance}. 
Several groups are now achieving coherent control manipulation of silicon spins. However, charge defects represent a potential challenge to forming predictable QD energetics and spin behavior. Inevitable charge defects in metal-oxide-semiconductor (MOS), for example, could strongly localize the electron even within a dot.
In this letter, we investigate the effect of fixed charges in the dielectric on the voltage tuned exchange (J) curve of a silicon MOS DQD. We investigate the impact of strong localizing Coulomb centers on the J-curve tunability accounting for the full silicon band structure.


We have developed an atomistic tight-binding (TB) based full configuration interaction (FCI) method to compute the multi-electron states of a DQD \cite{nemoci}. The single particle wavefunction is solved using the TB method for a DQD potential superimposed on the crystal potential. The QD wavefunctions from the TB solution form the single electron basis for the FCI. We combine the atomic scale effects with an exact many electron method to capture excited electron configurations, exchange, and correlation effects with unprecedented accuracy. The method enables simulation of realistic devices as millions of atoms can be simulated in high performance computing clusters \cite{Klimeck1}, and     
in general, the atomistic TB method captures realistic details of the devices including miscuts, step roughness, alloy disorder, valley splitting, confinement geometries, strain, and applied fields. 

\begin{figure}[htbp]
\center\epsfxsize=3.4in\epsfbox{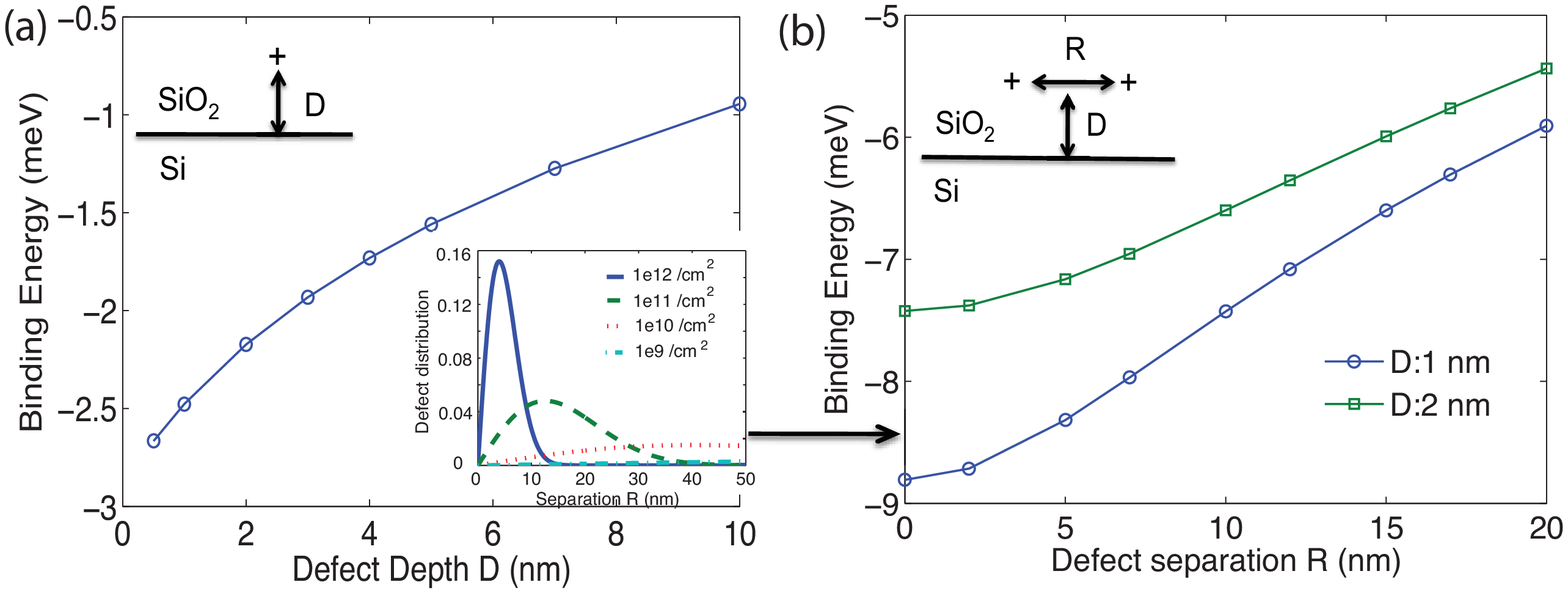}
\caption{Binding energies due to (a) a single charge defect as a function of depth $D$ into the oxide, and (b) two charge defects as a function of their separation $R$ for two different depths $D$. Inset of (a) shows the probability of finding two defects a distance $R$ apart for various defect densities. 
}
\label{fi1}
\end{figure}   

Fixed positive charges in the oxide near the Si-SiO$_2$ interface are common in MOS devices. The detailed chemical basis for the formation of these charge defects have been explored in Refs \cite{Defect1, Defect2}. Typical densities of these defects can range between $10^{9}$ /cm$^2$ to $10^{12}$ /cm$^2$ \cite{VLSI} and positive charge is a common polarity of the fixed defects. Each defect charge produces a Coulomb potential, the tail of which penetrates into silicon and forms a shallow potential well that can bind at least one electron below the conduction band (CB). 

Fig. 1(a) shows the binding energy of an isolated defect in SiO$_2$ as a function of the defect depth $D$, schematically shown in the top inset of Fig. 1(a), computed from the TB method. A single defect 
can bind an electron in silicon with energies of a few meV, comparable to the orbital energies of a QD. 
Inset of Fig. 1(a) shows the distribution of defect separation distances for various defect densities, as given by a Poisson distribution, whereas Fig. 1(b) shows the binding energies of two defects as a function of their lateral separation $R$ (inset of Fig. 1(b)) at a depth $D$ into the oxide. This shows that a cluster of defects can bind electrons even more strongly and are detrimental to QDs. The binding energies compare well with recent measurements of defect states using electron spin resonance techniques \cite{Ryan}. For these calculations, we used 
a large simulation domain of 100 nm $\times$ 100 nm $\times$ 50 nm to avoid finite size effects. 

The single electron TB Hamiltonian $h_i$ of a defect and a DQD, corresponding to the $i^{th}$ electron, is expressed as 
\begin{eqnarray} \nonumber
 h_i &=&  H_0+\textrm{min}\lbrace a(x-L)^2+\epsilon, a(x+L)^2\rbrace+ay^2+e F_z z \\
 && -\frac{e^2(1-Q)}{4 \pi \epsilon_{\textrm{Si}} \sqrt{(x-x_d)^2+(y-y_d)^2+(z+D)^2}}
 \label{eq1}
\end{eqnarray}
\noindent
where $H_0$ is the TB Hamiltonian of the host silicon formulated semi-empirically with the 10 band sp$^3$d$^5$s* model \cite{Klimeck1}. The second and the third terms describe the 2D parabolic potential energy of the QDs with curvature $a$ and center-to-center separation of $2L$. The left dot is subjected to a detuning of $\epsilon$ relative to the right dot. The 4th term is the potential energy due to a uniform vertical electric field $F_z$ that confines the electrons at the interface. The 5th term is the Coulomb potential energy of a defect located at $(x_d, y_d, -D)$ (interface being at $z=0$) along with its image charge correction given by the factor $Q=\frac{\epsilon_{\textrm{Si}}-\epsilon_{\textrm{SiO}_2}}{\epsilon_{\textrm{Si}}+\epsilon{\textrm{SiO}_2}}$ \cite{defect_V1}. We have used $\epsilon_{\textrm{Si}}=11.9 \epsilon_0$ and $\epsilon_{\textrm{SiO}_2}=3.9 \epsilon_0$ as the dielectric constants of Si and SiO$_2$ respectively, with $\epsilon_0$ as the permittivity of free space. Eq (\ref{eq1}) is defined in the silicon region only with $z \geq 0$. The Si-SiO$_2$ interface is modeled as a hydrogen passivated surface \cite{Lee}, as used in earlier works \cite{Klimeck2}. The full Hamiltonian is solved with a parallel Block Lanczos algorithm to extract the relevant eigenenergies and wavefunctions near the CB minimum using the Nanoelectronic Modeling Tool (NEMO 3D) \cite{Klimeck1}. The DQD system simulated in this work is comprised of about 1 million atoms, and was solved typically on 40 processors in 10 hours. 

Using a set of lowest energy single electron states obtained from the TB Hamiltonian, all possible antisymmteric two-electron configurations are constructed, and the two-electron Hamiltonian shown below is diagonalized in this basis,
 \begin{equation}
 H= h_1+ h_2+ \frac{e^2}{4 \pi \epsilon_{Si} |r_1-r_2|}
 \label{eq2}
\end{equation} 
\noindent 
where $h_1$ and $h_2$ are given by eq \ref{eq1}, and the 3rd term is the electron-electron repulsion term with electron coordinates $r_1$ and $r_2$. The solution of eq (\ref{eq2}) yields the two-electron states of the DQD-defect system. On average, we found about 12 single particle states corresponding to 66 two-electron configurations were needed for convergence.


\begin{figure}[htbp]
\center\epsfxsize=3.4in\epsfbox{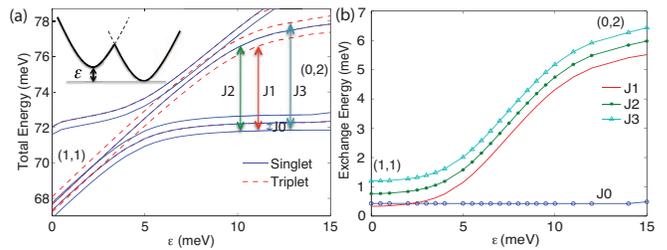}
\caption{ (a) Low lying energy states (total energies) of a double quantum dot without any charge defects as a function of the detuning energy $\epsilon$ indicate regions of $\epsilon$ where the DQD is more (1,1) or (0,2) in nature. The dot has curvature $a=0.0001$ eV/nm$^2$ equivalent to a 1D harmonic well with energy $E_0=8$ meV. The vertical field is $F_z=5$ MV/m. The energies are with respect to the CB minima of bulk silicon at $F_z=0$. Inset: 1D schematic of the detuned DQD potential. (b) Ideal (no defect) exchange (J) curves (i.e. various singlet-triplet splittings) of the DQD extracted from (a) as a function of $\epsilon$. 
}
\label{fi1}
\end{figure}   

In Fig. 2(a), we show the low energy two-electron spectrum (i.e. total energy of the two-electron Hamiltonian of eq (2)) of the silicon DQD as a function of the detuning energy $\epsilon$ (shown in the inset). At $\epsilon=0$, the DQD is in a (1,1) charge configuration with each electron in a separate dot. As $\epsilon$ is increased adiabatically, the DQD goes through a gradual charge transition to the (0,2) state. Since the wavefunction symmetries and the spatial extent change during this bias sweep, the exchange (J) energy varies as a function of $\epsilon$. It has been demonstrated experimentally in Ref \cite{Petta} for a GaAs DQD that the voltage controlled exchange can be used to provide a coherent rotation of the qubit encoded in the singlet-triplet basis.  

The bias dependence of the 2e spectrum is well-established for a GaAs DQD \cite{Petta, Taylor}. In Fig. 2, we have extended this calculation to a silicon DQD, where the multiple conduction band (CB) valleys add an additional degree of freedom to the electronic states. The TB calculations take into account all six CB valleys by representing the full bandstructure of silicon. Due to the applied vertical E-field of 5 MV/m in the simulations, the two k$_z$ valleys are split from the other four CB valleys. Interface and E-field induced inter-valley coupling causes a further valley splitting of about 0.1 meV between the two lowest k$_z$ valleys for the passivated interface used in this work. 

The salient feature of the two-electron energies, shown in Fig. 2(a), is the presence of a multiplet of levels which vary with the bias in a similar manner, whereas as a single level would be expected in a GaAs DQD. These multiplets are a result of the two k$_z$ valleys, as there exist multiple states with the same orbital envelope and spin wavefunctions, but with different rapidly oscillating Bloch wave components i.e. valley configurations. In the case of Fig. 2(a), each multiplet consists of three distinct lines offset by a roughly constant valley splitting. The exchange and Coulomb integrals used to evaluate the FCI Hamiltonian are strongly dependent on the valley configuration of the states \cite{Hada}, and methods that use a single valley approximation for silicon DQDs ignore potentially important details of the spin and valley physics. Since Fig. 2(a) is for B=0 T, the polarized triplets ($T_+$ and $T_-$) and the unpolarized triplet $T_0$ are degenerate, and only the $T_0$ triplets are shown. 

The exchange energy ($J$), defined as the splitting between the lowest singlet and triplet states, can assume multiple definitions here due to the valley degrees of freedom. In Fig. 2(a), we have labeled the energy difference of the lowest four triplets relative to ground state singlet, as $J0$, $J1$, $J2$ and $J3$ in order of increasing magnitude. Fig. 2(a) shows that there is a low lying triplet almost degenerate with the first excited singlet. These are states with predominantly orthogonal valley character \cite{Culcer, Hada} and hence small exchange splitting $J0$. The magnitude of $J0$ primarily depends on the valley splitting caused by the interface and the vertical E-field. The $J1$ splitting is due to the lowest triplet with similar valley configuration as the ground state singlet, and is analogous to the $J$-curve of a GaAs DQD. The magnitude of $J1$ mostly depends on the curvature of the dots, the magnitude of the tunnel barrier, and their separation distances. For large valley splitting, the $J0$ curve can be higher in energy than the $J1$ curve. Recent experiments have shown that the relaxation between triplets of different valley configuration can be strongly suppressed \cite{Rahman}, and hence our primary focus in this work will be the $J1$ splitting, although we will also show how a defect perturbs the $J0$ splitting.     

The four $J$-curves are plotted as a function of $\epsilon$ in Fig. 2(b). The $J0$ curve remains insensitive to detuning as the valley splitting is not affected by the lateral E-field. In the (1,1) charge configuration realized at small $\epsilon$, the high tunnel barrier between the dots reduces the overlap between the electronic wavefunctions. Hence the exchange energy $J1$ 
is small. As $\epsilon$ is increased the (1,1) and the (0,2) singlets anti-cross, and the later evolves into the ground state. The triplet, however, remains in the (1,1) configuration as the (0,2) triplet is at higher energy. In this regime, $J1$ increases steeply with $\epsilon$ as the singlet-triplet splitting increases (shown in Fig. 2(a) by the red arrow). At large enough $\epsilon$ the (1,1) and (0,2) triplets eventually anti-cross each other, and the lowest singlets and triplets are all in the (0,2) configuration. Since the (0,2) configuration is a high overlap system with the 2e confined in a single dot, the exchange energy is high. Any further detuning of the dots has only negligible effect on the (0,2) configuration, and $J1$ becomes insensitive to $\epsilon$. $J2$ and $J3$ behave in a similar manner to $J1$.

\begin{figure}[htbp]
\center\epsfxsize=3.4in\epsfbox{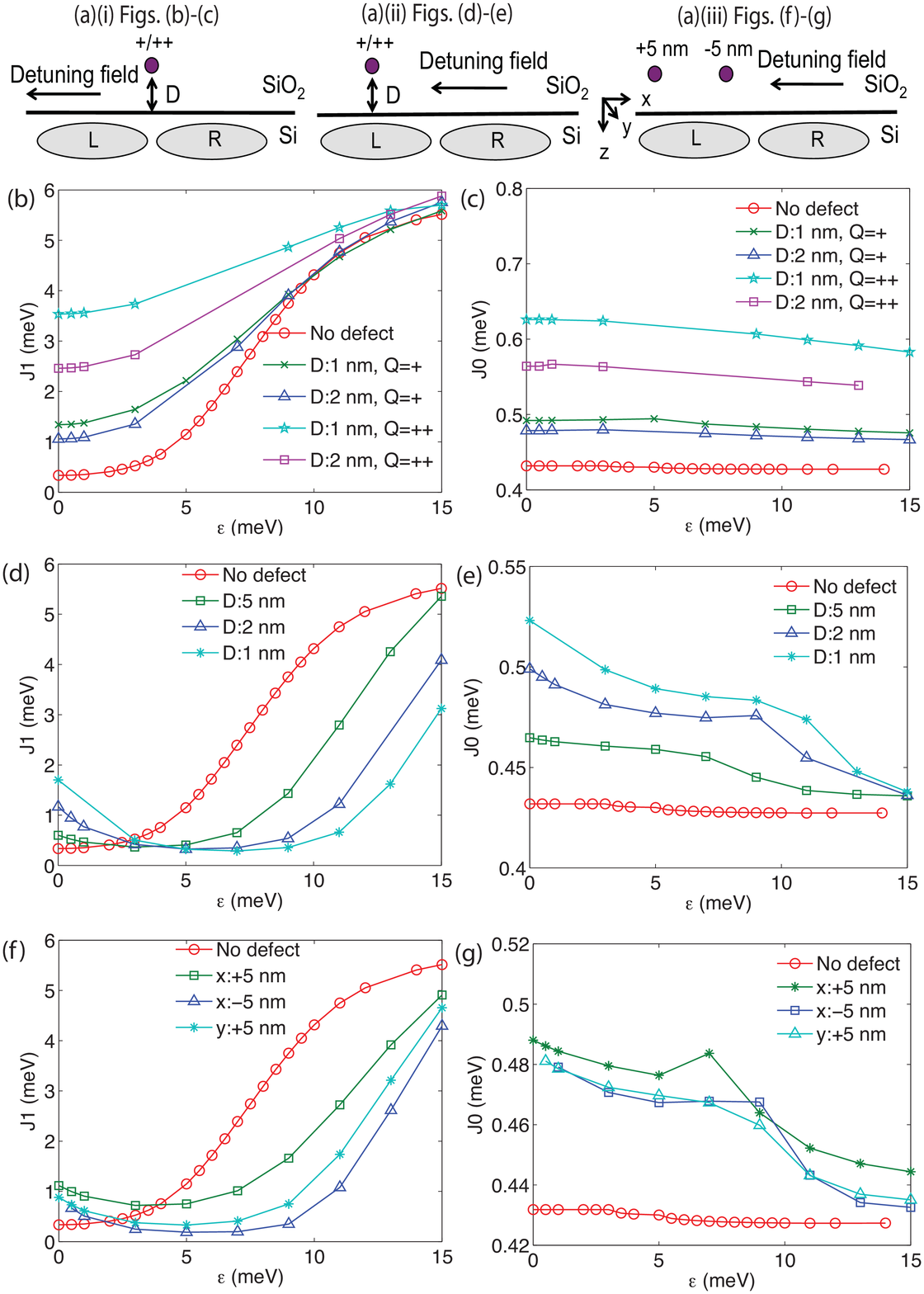}
\caption{(a) Schematic of a DQD labeled L and R with oxide charges (i) at the tunnel barrier, (ii) at the center of one dot, and (iii) at various locations relative to the center of one dot.  The corresponding exchange curves (b) J1 and (c) J0 of the DQD for a charge defect in the oxide at the tunnel barrier between the dots (case (a)(i)). (d) J1 and (e) J0 curves for an oxide defect at the center of the left dot (case(a)(ii)). (f) J1 and (g) J0 curves for different defect locations relative to the center of the dot (case (a)(iii)). 
}
\label{fi1}
\end{figure}   

Having established the voltage dependence of the $J$-curve in an ideal silicon DQD, we now analyze how the $J$-curve is perturbed in the presence of nearby oxide charges. In Fig. 3(b) and 3(c), we show the effect of a defect in the tunnel barrier between the dots (schematic of Fig. 3(a)(i)) on $J1$ and $J0$ respectively. A positive charge defect in the oxide lowers the potential barrier between the dots, and increases the overlap between the electronic wavefunctions in the (1,1) charge configuration. As a result, $J1$ increases at low detuning energy relative to the ideal $J1$-curve shown by the red dotted line. A lower barrier also makes the charge transition from (1,1) to (0,2) smoother, and the $J1$-curve flattens out in the transition region. As shown in Fig. 3(b), increasing the charge magnitude or decreasing the defect depth impacts the potential barrier more, as the corresponding $J1$-curves shift up and flatten out. Fig. 3(c) shows that the $J0$ curve typically shifts up if the influence of the defect increases. This is because a stronger defect potential enhances the vertical E-field, which increases the valley splitting in both dots.

As the positive defect charge is increased, the DQD is essentially transformed into a SQD insensitive to $\epsilon$. This means that if there are sufficient number of defects in close proximity to the tunnel barrier, the tunability of the $J1$-curve is hampered. In such a case, attempts to form a DQD could be futile as the electrons will always reside in a SQD charge configuration. 
Another consequence of a defect in the tunnel barrier is that it could result in an always `on' exchange gate, as the exchange interaction in the (1,1) occupation of a DQD is increased. 

In Fig. 3(d) and 3(e), we show the $J$-curves for a defect located at the center of the left dot in the xy-plane with various depths into the oxide (schematic of Fig. 3(a)(ii)). In this case, the defect lowers the potential of the left dot relative to the right even at $\epsilon=0$, producing a natural detuning bias for the dots. The magnitude of this detuning increases as the defect is located closer to the dot (i.e. depth $D$ decreases). Fig. 3(d) shows that the $J1$-curves are translated more in $\epsilon$ as $D$ decreases, as more detuning bias needs to be applied to compensate for the defect induced detuning and to restore the electrons to the (1,1) charge configuration. Since the defect has only a small effect on the tunnel barrier, the slope of the J-curves remain the same, unlike the cases considered in Fig. 3(b). The presence of a defect in one dot relative to the other also causes an asymmetry in the J-curve between the (0,2) and the (2,0) configurations as the defect may produce a stronger confinement for the electrons and may result in a stronger exchange splitting. The $J0$-curves of Fig. 3(e) exhibit a slope with $\epsilon$ due to the fact that the defect produces a larger valley-splitting in the left dot compared to the right. Hence the $J0$ curve decreases in magnitude as the electrons go from a (2,0) to a (0,2) transition.

In Fig. 3(f) and 3(g), we show the effect of a defect at various locations relative to the dot (schematic of Fig. 3(a)(iii)). The defect depth $D$ is fixed at 2 nm, and the dot centers are 20nm apart along the x-axis. The label $x=+5$ nm indicates a defect in between the left dot and the tunnel barrier, displaced 5 nm from the left dot $L$ in the x-direction. Similarly, the label $x=-5$ nm indicates a defect located 5 nm right of the center of the dot and 15 nm from the tunnel barrier. The $J$-curves corresponding to these two cases can be understood as a combination of Fig. 3(b)-(e). The defect essentially detunes the left dot and thus translates the $J1$-curve, but to a lesser extent than it does in Fig. 3(c) where the defect is at the dot center. A defect in between the dot and the tunnel barrier, in addition, lowers the tunnel barrier, which manifests as a flatter $J1$-curve. The label $y=+5$ indicates a defect displaced 5 nm from the dot center perpendicular to the DQD axis. The $J1$-curve in this case is also translated, and lies between the two previous curves, as expected. Similarly, the $J0$-curves also exhibit an increased magnitude and a slope as expected.


We have done a full configuration interaction calculation of a Si DQD using TB wavefunctions that account for valleys and atomistic details. Our results show that charge defects in the oxide near a silicon DQD can determine the characteristics of the multi-valley exchange-curve. Defects in the tunnel barrier affect the $J$-curve most as they can hamper the tunability of a SQD to a DQD. Defects at other locations mainly manifest as a translation in the $J$-curve to a different detuning bias. The combined TB and CI method developed to perform these calculations represent a significant advancement in the computational simulation methods for nanostructures, as it integrates the atomic scale details with an exact many-electron theory, and can easily scale up to million atom systems.

\begin{acknowledgments}
Sandia is a multiprogram laboratory operated by Sandia Corporation, a Lockheed Martin Company, for the United States Department of Energy's National Nuclear Security Administration under Contract No. DE-AC04-94AL85000. RR acknowledges Gerhard Klimeck for the NEMO 3D code, and W. Witzel, R. Young, and X. Gao for insightful discussions.
\end{acknowledgments}

Electronic address: rrahman@sandia.gov

\end{document}